\documentclass[preprint,12pt]{elsarticle}

\usepackage{graphicx}

\usepackage{amssymb}

\usepackage[nodots,nocompress]{numcompress}

\usepackage{lineno}

\usepackage{textcomp} 

\usepackage{epstopdf}

\biboptions{super}

\journal{Nucl. Instrum. Methods A}

\def\be{\begin{equation}}
\def\ee{\end{equation}}
\def\bea{\begin{eqnarray}}
\def\eea{\end{eqnarray}}

\begin{document}

\newcommand{\mnv}{\mathrm{MINER}{\nu}\mathrm{A}}
\newcommand{\trp}{TriP-t\,}

\newcommand{\Rutgers}{Rutgers, The State University of New Jersey, Piscataway, New Jersey 08854, USA}
\newcommand{\Hampton}{Hampton University, Dept. of Physics, Hampton, VA 23668, USA}
\newcommand{\Dortmund}{Institute of Physics, Dortmund University, 44221, Germany }
\newcommand{\Otterbein}{Otterbein College,1 South Grove Street, Westerville, OH, 43081,USA}
\newcommand{\JMU}{James Madison University, Harrisonburg, Virginia 22807, USA}
\newcommand{\Florida}{Department of Physics, University of Florida, Gainesville, FL 32611,USA}
\newcommand{\UCIrvine}{Department of Physics and Astronomy, University of California -- Irvine, Irvine, California 92697-4575, USA}
\newcommand{\CBPF}{Centro Brasileiro de Pesquisas F\'{i}sicas, Rua Dr. Xavier Sigaud 150, Urca, Rio de Janeiro, RJ, 22290-180, Brazil}
\newcommand{\INRM}{Institute for Nuclear Research of the Russian Academy of Sciences, 117312 Moscow, Russia}
\newcommand{\Jlab}{Jefferson Lab, 12000 Jefferson Avenue, Newport News, VA 23606, USA}
\newcommand{\Pittsburgh}{Department of Physics and Astronomy, University of Pittsburgh, Pittsburgh, Pennsylvania 15260, USA}
\newcommand{\Guanajuato}{Universidad de Guanajuato. Lascura\'{i}n de Retana No. 5. Col. Centro. Guanajuato 37150, Guanajuato. M\'{e}xico}
\newcommand{\Athens}{Department of Physics, University of Athens, GR-15771 Athens, Greece}
\newcommand{\Tufts}{Physics Department, Tufts University, Medford, Massachusetts 02155, USA}
\newcommand{\WM}{Department of Physics, College of William \& Mary, Williamsburg, Virginia 23187, USA}
\newcommand{\FNAL}{Fermi National Accelerator Laboratory, Batavia, Illinois 60510, USA}
\newcommand{\Purdue}{Department of Chemistry and Physics, Purdue University Calumet, Hammond, Indiana 46323, USA}
\newcommand{\MCLA}{Massachusetts College of Liberal Arts, 375 Church Street, North Adams, MA 01247, USA}
\newcommand{\UMD}{Department of Physics, University of Minnesota -- Duluth, Duluth, Minnesota 55812, USA}
\newcommand{\Northwestern}{Northwestern University, Evanston, Illinois 60208, USA}
\newcommand{\UNI}{Universidad Nacional de Ingenier\'{i}a, Tupac Amaru Avenue 210, Lima, Per\'u}
\newcommand{\agagoOverride}{Secci\'on F\'{\i}sica, Departamento de Ciencias, Pontificia Universidad Cat\'olica del Per\'u, Apartado 1761, Lima, Per\'u }
\newcommand{\Rochester}{University of Rochester, Rochester, New York 14610 USA}
\newcommand{\Austin}{Department of Physics, University of Texas, 1 University Station, Austin, Texas 78712, USA}
\newcommand{\USM}{Departamento de F\'isica, Universidad T\'ecnica Federico Santa Mar\'ia, Avda. Espa\~na 1680 Casilla 110-V Valpara\'iso, Chile}
\newcommand{\schulteOverride}{Temple University, Philadelphia, PA 19122-6099, USA}
\newcommand{\bradfordOverride}{Argonne National Laboratory, 9700 S. Cass Avenue, Lemont, IL 60439, USA}

\begin{frontmatter}



\title{The MINER$\nu$A Data Acquisition System and Infrastructure}


  \author[Rochester]    {G.~N.~Perdue\corref{cor1}}
  \cortext[cor1]{Corresponding Author}
  \ead{perdue@fnal.gov}
  \author[FNAL]         {L.~Bagby}
  \author[FNAL]         {B.~Baldin}
  \author[FNAL]         {C.~Gingu}
  \author[FNAL]         {J.~Olsen}
  \author[FNAL]         {P.~Rubinov}
  \author[Rutgers]      {E.~C.~Schulte\fnref{fnref1}}
  \fntext[fnref1]{Present Address: \schulteOverride}

  \author[Rochester]    {R.~Bradford\fnref{fnref2}}
  \fntext[fnref2]{Present Address: \bradfordOverride}
  \author[USM]          {W.~K.~Brooks}
  \author[CBPF]         {D.~A.~M.~Caicedo}
  \author[CBPF]         {C.~M.~Castromonte}
  \author[Rochester]    {J.~Chvojka}
  \author[CBPF]         {H.~da~Motta}
  \author[Pittsburgh]   {I.~Danko}
  \author[WM]           {J.~Devan}
  \author[Pittsburgh]   {B.~Eberly}
  \author[Guanajuato]   {J.~Felix}
  \author[Northwestern] {L.~Fields}
  \author[CBPF]         {G.~A.~Fiorentini}
  \author[agagoOverride]{A.~M.~Gago}
  \author[UMD]          {R.~Gran}
  \author[FNAL]         {D.~A.~Harris}
  \author[CBPF,UNI]     {K.~Hurtado}
  \author[Rochester]    {H.~Lee}
  \author[MCLA]         {E.~Maher}
  \author[Rochester]    {S.~Manly}
  \author[Rochester]    {C.~M.~Marshall}
  \author[Rochester]    {K.~S.~McFarland}
  \author[Rochester]    {A.~Mislivec}
  \author[Florida]      {J.~Mousseau}
  \author[Florida]      {B.~Osmanov}
  \author[FNAL]         {J.~Osta}
  \author[Pittsburgh]   {V.~Paolone}
  \author[Rutgers]      {R.~D.~Ransome}
  \author[Florida]      {H.~Ray}
  \author[FNAL]         {D.~W.~Schmitz}
  \author[UCIrvine]     {C.~Simon}
  \author[UNI]          {C.~J.~Solano~Salinas}
  \author[Rutgers]      {B.~G.~Tice}
  \author[Hampton]      {T.~Walton}
  \author[Rochester]    {J.~Wolcott}
  \author[WM]           {D.~Zhang}
  \author[UCIrvine]     {B.~P.~Ziemer}
  \address[CBPF]{\CBPF}
  \address[WM]{\WM}
  \address[FNAL]{\FNAL}
  \address[Hampton]{\Hampton}
  \address[MCLA]{\MCLA}
  \address[Northwestern]{\Northwestern}
  \address[agagoOverride]{\agagoOverride}
  \address[Rutgers]{\Rutgers}
  \address[Guanajuato]{\Guanajuato}
  \address[UNI]{\UNI}
  \address[USM]{\USM}
  \address[UCIrvine]{\UCIrvine}
  \address[Florida]{\Florida}
  \address[Pittsburgh]{\Pittsburgh}
  \address[UMD]{\UMD}
  \address[Rochester]{\Rochester}

\begin{abstract}
MINER$\nu$A (Main INjector ExpeRiment $\nu$-A) is a new few-GeV neutrino cross section experiment that began taking data in the FNAL NuMI (Fermi National Accelerator Laboratory Neutrinos at the Main Injector) beam-line in March of 2010.  MINER$\nu$A employs a fine-grained scintillator detector capable of complete kinematic characterization of neutrino interactions. This paper describes the MINER$\nu$A data acquisition system (DAQ) including the read-out electronics, software, and computing architecture.   
\end{abstract}


\end{frontmatter}

\linenumbers

\section{Introduction}
\label{sec:intro}
$\mnv$ (Main INjector ExpeRiment $\nu$-A) is a low-to-medium energy neutrino-nucleus scattering experiment  at FNAL (Fermi National Accelerator Laboratory).  
It utilizes the high-intensity NuMI (Neutrinos at the Main Injector) beamline and sits in the MINOS (Main Injector Neutrino Oscillation Search) Near Detector Hall, directly in front of the MINOS Near Detector and 105 meters below grade.  
$\mnv$ does not extend far enough in the beam direction to range out muons with momentum over about 2 GeV/c, so the MINOS Near Detector is directly employed as a muon spectrometer. 
The references contain more detailed information on the physics program and general details of $\mnv$ \cite{refmnvcdr}, MINOS \cite{refminos}, and the NuMI beam \cite{refnumi}.

This document discusses the $\mnv$ data acquisition system (DAQ) in two steps.  
First, the custom electronics are described, followed by a description of the DAQ software and computing architecture.  

The initial specifications for the DAQ readout system asked for a readout rate of $\sim$100 kB/s with a duty factor defined by one $\sim 10$ microsecond beam spill every 2.2 seconds \cite{refmnvtdr}. Fortunately, those specifications were surpassed as our initial estimate for required data throughput was low by about a factor of five. 

\section{Overview of the Readout Scheme in $\mnv$}
\label{sec:overview}
The DAQ, slow controls (SC), and near-online monitoring (``nearline'') systems are built around $\mnv$-specific readout electronics.  
$\mnv$ utilizes plastic scintillator as its fundamental detector technology.  
Light is delivered via wavelength shifting fiber to Hamamatsu R7600 64-channel multi-anode photomultiplier tubes (PMTs) \cite{refm64}.  
A small number of single-anode PMTs are also mounted for a ``Veto Wall'' detector that sits in front of the main $\mnv$ detector.

Each PMT is read out by a Front End Board (FEB) \cite{reffeb} mounting six Application-Specific Integrated Circuit (ASIC) chips (\trp chips) \cite{Rubinov:2005zz} that digitize and store charge using pipeline ADCs.
Input charges from the PMT anodes are divided into high, medium, and low gain channels using a capacitive divider to increase the dynamic range.
The high gain is 1.25 fC/ADC, the medium is 4 fC/ADC, and the low is 15.6 fC/ADC.
The FEBs generate the high voltage for most of the PMTs using an on-board Cockroft-Walton (CW) generator. 
The control circuitry resides on the FEB, while the CW chain itself resides on the base, with appropriate taps for each dynode.
A small number of the PMTs are single-anode and use resistor-base technology for voltage control.
These PMTs are used to read out the Veto Wall and employ a separate voltage control system discussed below.
All FEB operations are controlled by a Spartan 3E Field-Programmable Gate Array (FPGA) chip. 
The FPGAs decode timing signals received over the unshielded twisted pair (UTP) cables, sequence the \trp chips and decode and respond appropriately to communication frames received over the data link. The FPGA also controls the CW and other aspects of FEB operation.
For data collection, the boards are daisy-chained together (into ``chains'') using standard UTP ethernet networking cables with a custom protocol and Low Voltage Differential Signaling (LVDS). Of the four pairs in the cable, one is dedicated to timing, including clock and encoded signals, one is dedicated for data, one is used to indicate the sync-lock status of the data Serializer/Deserializer (SERDES) and one for a test pulse. 

The readout chain is connected  at both ends to a custom VME module called the Chain ReadOut Controller (CROC) \cite{refcroccrim}.  
A CROC supports four Front End (FE) channels, each serving one chain of up to fifteen FEBs, though no chain is longer than ten FEBs in $\mnv$.  
Each of the four CROC channels contains a 6 kB dual-port memory for storing messages (called ``frames'') from the FEBs.  
The CROCs in turn receive timing and trigger commands from another custom VME module, the CROC Interface Module (CRIM) \cite{refcroccrim}, each of which distributes timing to up to four CROCs.  
$\mnv$ uses two VME crates, each containing a CAEN V2718 crate controller \cite{refcaen2718}, two CRIMs, and roughly a half-dozen CROCs.  
There is no hit-based trigger but rather a timing-based integration gate synchronized to the FNAL Main Injector (MI) timing signals.  
Finally, the CRIMs receive information from the timing system of the MINOS experiment (used for event matching) and the state of the MI via the $\mnv$ Timing Module (MvTM) \cite{refmvtm}.  
The MvTM does not have a VME interface, but uses the VME crate for power.
The start time in MINOS detector timing coordinates is logged as a 28-bit number in the DAQ Header bank data.  
See Table \ref{tbl:hwinst} for a complete module and card accounting.
Only one MvTM is required to service all timing needs.  
Each individual crate is readout by a computer (PC) mounting a CAEN A2818 PCI card \cite{refcaen2818} that communicates with the crate controller module via optical cable.

\begin{table}[htb]
\begin{center}
\begin{tabular}{|c|c|c|c|c|}
\hline
Era & VME Crates & CROCs & FEBs & Total Channels \\
\hline
\hline
Fall 2009 & 2 & 8 & 272 & 17,408 \\
\hline
Spring 2010 & 2 & 14 & 491 & 31,424 \\
\hline
Summer 2010 + & 2 & 15 & 509 & 32,576 \\
\hline
\end{tabular}
\end{center}
\caption{
$\mnv$ read-out hardware installations.}
\label{tbl:hwinst}
\end{table}

An integration gate on the FEBs is opened synchronously with the delivery of neutrino beam spills.  
Beam spills delivered by the MI are approximately 10 microseconds long, and are delivered every 2.2 or 2.06 seconds, depending on the MI operating mode.  
Our readout gate opens 500 nanoseconds before the arrival of the spill and remains open 5.5 microseconds after the end.  
During normal data-taking, one additional calibration gate is recorded between neutrino spills, either a random sampling of the detector noise (a pedestal gate) or in coincidence with LEDs flashing the PMTs (a ``light-injection'' gate).
Pedestals are used to establish baseline readout levels and light-injection to monitor gain drifts in the PMTs by studying the photo-statistics of the LED signals.

At the end of the gate, the readout system runs a software loop over all of the FEBs and collects data in the form of device ``frames'' (well-defined and formatted byte data with attached headers) one at a time for the state of the high-voltage, hit timing, and finally the hit blocks themselves.  
Each frame is passed through a chain to a CROC FE channel where it is stored briefly before being passed to a readout computer for archival and monitoring processes.
  
Readout of the electronics is done in pairs of nested loops. 
Requests for data are sent to the first FEB on each FE channel before looping back to retrieve the data from the CROC memory. 
Then the next set of FEBs are handled in another dual loop over FE channels, etc. This allows the DAQ computers to communicate with the next CROC while the previous one is fetching data from the FEBs on one of its readout chains.

In normal data-taking $\mnv$ reads out in a ``zero suppression'' mode. 
In this mode, the DAQ first reads the discriminator frame data for each board and stores frames that contain hits above the discriminator threshold (about 70 fC).    
It then loops over those boards and reads the FPGA registers and all time-stamped ADC hit block frames.  
The FEBs are  configured to enable readout of up to eight ADC blocks: seven time-stamped hits and one ``un-timed'' hit at the end of the integration gate.
The un-timed hit carries no discriminator timing information and contains all the integrated charge between the last timed hit and the end of the gate.
The FPGA registers contain configuration data, high-voltage read-back, and the local ``gate time stamp'' - the time the gate was  opened for integration in the local FEB counter time coordinates.
Boards without hits above threshold are not read out in this loop and neither are un-time-stamped ADC frames (charge collected below threshold and stored until the end of the integration gate). All seven available time-stamped hits are read out.
During construction and  early low-energy configuration running, a configuration with five time-stamped hits and one ``un-timed'' end-of-gate hit was used.
See Table \ref{tbl:framesizes} for a listing of the sizes of each data type.

\begin{table}[htb]
\begin{center}
\begin{tabular}{|c|c|l|}
\hline
Frame Type & Size (bytes) & Comments \\
\hline
\hline
FPGA Programming  & 66 & Spartan 3E control registers.  \\
\hline
ADC  & 444 & Hit data.  \\
\hline
Discriminator & 16 + 40 per & Time stamps for each hit. \\
 & \trp per hit  & \\
\hline
TriP Programming & 652 (read)  & \trp internal registers. \\
 &  758 (write) & \\
\hline
DAQ Header & 48 & Not a true ``frame.'' \\ 
 & & $\rightarrow$ End of gate marker. \\
\hline
\end{tabular}
\end{center}
\caption{
Frame sizes per FEB for $\mnv$ data objects.  
The FPGA programming frames track configuration details and provide data on the high voltage, gate initiation time, and some useful error-checking bits.  
The \trp chips were originally designed for the D0 experiment \cite{refdzero} and so carried timing data that is not usable by $\mnv$.  These data were trimmed in the packing algorithm for more efficient use of space.  This reduced storage bloat, but did not meaningfully impact speed. 
Only discriminator frames are of variable size.  
The \trp programming frames are not read or written to during data acquisition. They are only accessed during configuration stages by the slow controls.  }
\label{tbl:framesizes}
\end{table}

Despite using more than 32,000 active channels in $\mnv$, the quantity of data per neutrino spill is not large due to low interaction cross-sections.  
With a 2.2 second MI cycle and $35 \times 10^{12}$ protons on target per spill, roughly 1 MB per spill for the entire detector in zero-suppression mode running in the NuMI low-energy configuration is collected, where the most frequent neutrino energy is $\sim$3 GeV. 
About half of those data are beam spill data and half are calibration data.  

Readout time for a single frame is approximately 500 microseconds. 
Most of that time is in message preparation and in setting up a block transfer; for our data the readout speeds per frame are largely insensitive to the size of the frame.
Physics gate readouts typically require 1200 frames on average, while calibration gate readouts require 1600 (channel occupancy is lower in physics gates). 
Therefore, reading a typical ``cycle'' (single physics gate plus single calibration gate) requires 1.4 seconds. 
Calibration readout times are very stable, but physics gate readout times vary with the level of activity in the detector. 
The DAQ is programmed to protect the integrity of physics gates above all else. 
It will only attempt a calibration readout in the case where the physics readout is accomplished in under 0.9 seconds.
If the subsequent calibration readout extends beyond 0.9 seconds, the DAQ will dump that readout and simply arm for the next physics spill.
However, the DAQ will not interrupt readout of a physics gate. 
Propagation of the start gate signal is blocked in hardware and only unlatched after readout is complete.

\section{System-Wide Timing Signals}
\label{sec:systwidetiming}
\begin{figure}[hbtp]
  \centering 
  \includegraphics[width=0.7\textwidth]{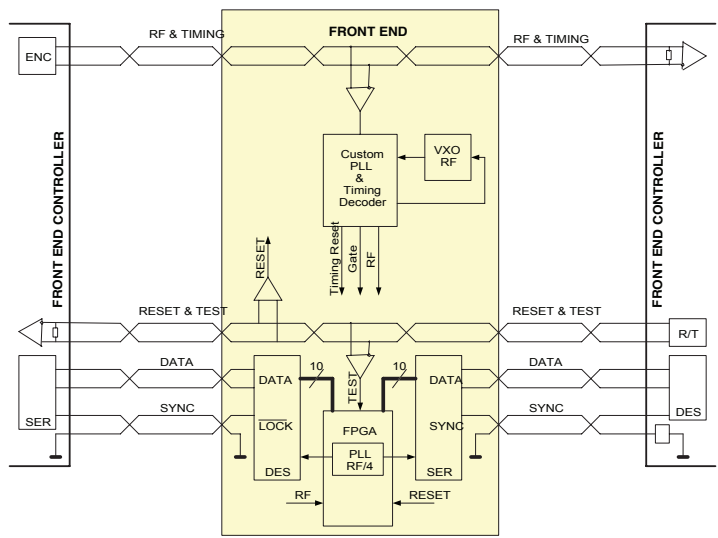}
  \caption{The DAQ Loop Timing Signals. \cite{refcroc}}
  \label{fig:daqloopsignals}
\end{figure}

Time-stamps are recorded for the $\mnv$ experiment using a single 53.1 MHz crystal on the MvTM which is distributed to all of the electronics components - from MvTM to the CRIMs, from the CRIM to the CROCs, from the CROCs to the FEBs. See Fig. \ref{fig:daqloopsignals} for an illustration of the timing signals loop. 
Each board regenerates the clock using phase-locked loop (PLL) circuits. 
The 53.1 MHz clock is distributed on the readout cable, but the FEBs use the onboard PLL to multiply the frequency by two. 
All times in $\mnv$ are defined in units of system ticks, or clock ticks.  A clock tick is equal to
\begin{equation}
	1/f = 1/(2 \times 53.1 \times 10^6 \,\mathrm{Hz}) = 9.4 \,\mathrm{ns}
\end{equation}
Additionally the FPGA on the FEB uses a quadrature circuit, providing a granularity of $\sim$2.4 ns for the discriminator time stamps.
Throughout the remainder of this document, the term ``clock tick'' will refer to the unit used by the FEBs ($\sim$9.4 ns) unless explicitly noted otherwise.  

The FEBs, CROCs and CRIMs can operate using on board oscillators, but FEBs and
CROCs in $\mnv$ are always operated  by synchronizing their clocks to a radio frequency (RF) signal provided by the CRIMs. 
The CRIMs are typically operated in  ``MvTM Mode,'' and locked to the MvTM clock, but in certain situations are placed in ``Internal Mode,'' where the clock is supplied by the internal crystal oscillator.
See Table \ref{tbl:crimtiming}.
$\mnv$ electronics were designed to operate without a trigger, so the cosmic and Test-Beam modes are special.  
In those cases, a coincidence between a live gate ($\approx 16\,\mu$s wide) and an external logic pulse is searched for in order to make readout decisions.  
Because of dead-time between gates, the live-time is low in those operating modes ($\sim 10\%$).

\begin{table}[htb]
\begin{center}
\begin{tabular}{|c|c|}
\hline
Data-Taking Mode & CRIM Clock Mode \\
\hline
\hline
Pedestal & Internal  \\ 
\hline
Cosmic Rays & Internal \\
\hline
FTBF (Test-Beam) Beam & Internal \\
\hline
NuMI Beam & MvTM \\
\hline
Light Injection (LI) & MvTM \\
\hline
Mixed Mode (Beam with LI or Pedestal) & MvTM \\
\hline
\end{tabular}
\end{center}
\caption{CRIM timing modes for different data-taking modes.  ``FTBF'' refers to the Fermilab Test Beam Facility.  }
\label{tbl:crimtiming}
\end{table}

When operating in the internal timing mode, each CRIM supplies its own clock and either the DAQ system sends a new open-gate command after each successful readout or the CRIM opens the gates independently at a frequency set by the user ranging between 0.5 Hz and 52 kHz.  
The first sort of internal clock mode operation is used for standalone pedestal runs and the second requires an external trigger in coincidence to initiate readout (cosmic or test beam).

\begin{figure}[hbtp]
  \centering 
  \includegraphics[width=0.7\textwidth]{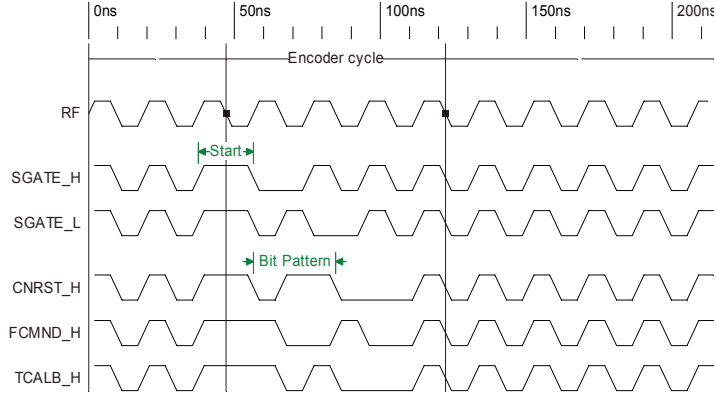}
  \caption{The Timing Encoder Signals. \cite{refcroc}}
  \label{fig:timingencodersignals}
\end{figure}

Encoded clock and trigger signals enter the CROCs over LVDS via RJ-45 input connectors:
\begin{enumerate}
\item RF - Radio Frequency
\item SGATE - Start Gate (open an integration gate)
\item CNRST - Counter Reset
\item FCMND - Fast Command (special commands)
\item TCALB - Timing Calibration Pulse
\end{enumerate}
See Fig. \ref{fig:timingencodersignals} for the structure of the encoded commands on the clock line.

In either Internal or MvTM timing mode, the CRIM can receive external signals via a set of LEMO connectors on its front panel or via an RJ-45 connector.  
The MvTM is able to recieve timing signals from the MINOS Master Timing Module over UTP Cat5e using LVDS.
The MvTM additionally can send and relay Low Voltage Transistor-Transistor Logic (LVTTL) signals through the pairs of in-and-out LEMO connectors.
One LEMO input is used to receieve a LVTTL signal providing MI cycle status from the Accelerator Division (this signal ultimately detemines the start-gate time). 
The other two LEMO connectors are used to fire the light injection system (discussed later).

\begin{figure}[hbtp]
  \centering 
  \includegraphics[width=0.7\textwidth]{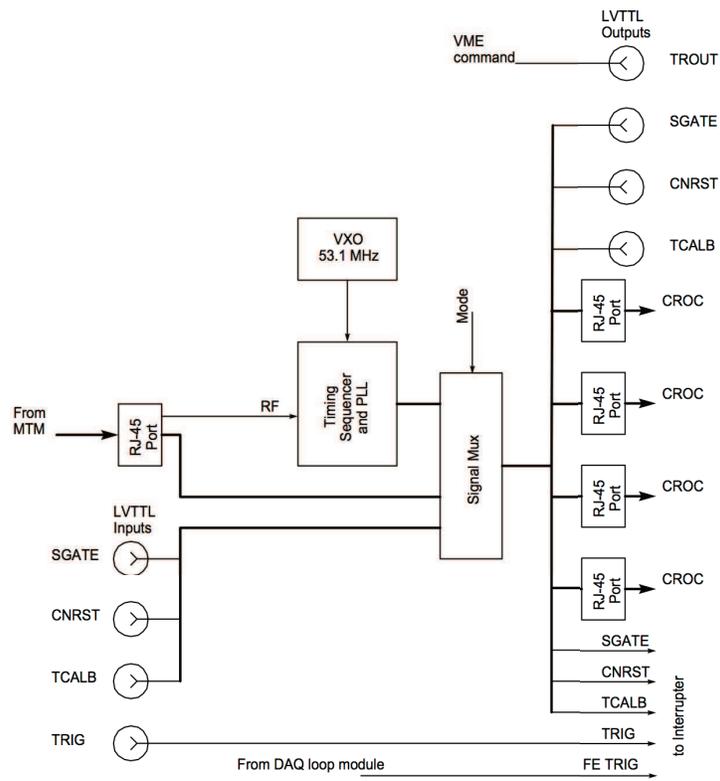}
  \caption{The CRIM (timing module) block diagram. \cite{refcrim}}
  \label{fig:timingmodule}
\end{figure}

During normal operation, the CRIM clock is derived from the MvTM. 
In principle, the MvTM can sync to the MINOS clock signal but it runs free against that clock and  the MvTM internal oscillator is used to synchronize all of the clocks.
The arrival time of the MINOS start-gate signal is recorded in a register in the CRIM, however. 
The MINOS and $\mnv$ clocks are therefore running in sync with an unknown phase shift.
When in MvTM mode, the CRIM will raise an interrupt when an external trigger pulse is detected - for example, when the spill signal comes from the MI.
This interrupt begins the readout sequence and blocks until it is reset after readout is complete.
Fig. \ref{fig:timingmodule} illustrates the timing module logic.

The CRIM interprets Start-gate (SGATE), trigger (TRIG), timing calibration (TCALB), and counter-reset (CNRST) LVTTL signals and generates corresponding output in a slightly different way for each of the different timing modes.
\begin{enumerate}
\item SGATE is sent through SGATE OUT in all timing modes.  
\item TRIG is sent only in MvTM mode, and only in case of a software command to open a gate.
\item TCALB is sent synchronously with SGATE in most CRIM firmware versions.  
It can be configured to only send this signal when an  external TRIG is supplied.
\item CNRST is sent synchronously with SGATE in most CRIM firmware versions.  
The FEBs interpret this as a ``Counter Load'' signal and set their counters to a value specified in the FPGA programming registers.  
This is part of the timing synchronization scheme for the FEBs.
\end{enumerate}
The clock and data travel only in one direction over the readout chains- from the CROC to each of the FEBs in turn and arriving back at the CROC. The CROC checks that the clock is present at the receiving end and will set an error flag if it is not. Further data acquisition is not possible until this error is cleared.

Our configuration for data-taking in the NuMI hall is:
\begin{enumerate}
\item A LEMO cable from the accelerator division rack is attached to LEMO input 1 on the MvTM.
A signal synchronized with the MI operational cycle arrives $\sim$215 $\mu$s before a spill.
A delay is programmed into the FEB FPGA's to account for the offset.  
This signal is sent to the CRIM from the MvTM via a CAT3 cable, and propagated from there through the CROCs to the FEBs.  
The FEBs use only the rising edge of the SGATE signal to open a gate.  
The falling edge is ignored.
\item A dedicated CAT3 cable is routed from the MINOS clock rack into the Master Clock Interface connector on the MvTM. 
This cable carries CNRST, TCALB, and SGATE encoded clock signals. The CRIM receives those signals as:
\begin{itemize}
\item $\mnv$ CNRST - MINOS CNRST. Set by the MINOS GPS clock to reset asynchronous to the other signals at a rate of  1 Hz.  
This signal is currently masked from the FEBs, but used to reset the counter on the CRIM.  
\item $\mnv$ TCALB - The MINOS SGATE signal is mapped here in the MvTM.  
The clock time is recorded (using on-board counters) for the arrival of this signal and stored on a register in the CRIM that is read at the end of the acquisition cycle.
\item $\mnv$ SGATE - This receiver is masked within the MvTM and the CRIM never sees a signal of this type.
\end{itemize}
\item All of the CRIMs are daisy-chained together via the TRIG port, going out from the ``Master CRIM'' (the CRIM with the lowest address in the VME crate with the lowest address) and then in and out of all subsequent CRIMs (incrementing by CRIM address before crate address) until the input on the last CRIM (the CRIM with the highest address in the VME crate with the highest address) is reached.
This chain is responsible for propagating the start signal in light injection data taking.  
A software command is sent to the Master CRIM only and propagated to the others via this electrical connection.  
Additionally a signal is sent from the Master CRIM SGATE connector to the light injection box.  
This output is passed through the MvTM itself for shaping and delay.
\item The CRIM is run in one of two timing modes (see Table \ref{tbl:crimtiming}).  
In the Internal and MvTM modes a sequencer that sends SGATE, TRIG, and CNRST to the FEBs is employed.  
If running in ``cosmic'' mode for the $\mnv$ Test Beam Experiment,  TCALB is additionally sent if an interrupt was raised.  
\end{enumerate}

\section{Hit Timing}
\label{sec:hittiming}
The $\mnv$ FEBs mount six \trp chips.  Four are dedicated to the high and medium gain and two are dedicated to the low gain.  
Due to this division, the \trp chips must be grouped together when storing hits in order to avoid some potential ambiguities.  
This has important consequences for hit timing.

\subsection{\trp ``Push'' Behavior}
\label{sec:pushbehavior}
Understanding hit storage is important for understanding hit timing. The \trp has a 48-cell deep analog pipeline and integrated charge is pushed into this pipeline when the discriminator fires according to a process described below. 
Each cell of charge corresponds to a ``hit.''
This behavior is not internal to the \trp chip, but is controlled but the FPGA firmware.
The readout depth (number of total allowed hits) is limited by the internal memory of the FPGA and the way digitized data from the hits is stored in the FPGA.
There is flexibility in the maximum number of hits that may be stored, but in standard configurations prior to 2012, $\mnv$ stored up to eight hits. 

Charge is integrated over a time window set by the user.  
The whole window is referred to as a ``gate.''  
Activity is separated in time inside this gate by latching the system counter when the integrated charge goes over a user-defined threshold and then storing the charge integrated up to that time (this process is called ``pushing the pipeline'').  
After storing the hit, the integrators are reset and re-opened.  
At the end of the gate, the analog information stored in the pipeline is digitized into RAM with one ADC block (or ``hit'') for each stored set of charges in the pipeline (for each ``pipe'').  
The different times associated with these processes are tunable by the user.  
In $\mnv$, the standard gate length is 1702 clock ticks ($\approx$16 $\mu$s).  
The integration time after a discriminator fires is 16 clock ticks ($\approx$150 ns), and the reset  requires 20 clock ticks ($\approx$188 ns).

The six \trp chips on each FEB are numbered 0-5, and each carries 32 channels.  
\trp chips 0-3 each operate 16 channels in the high gain mode and 16 channels in the medium gain mode. 
\trp chips 4 \& 5 each operate 32 channels in the low gain mode.  
There are 64 channels on each multi-anode PMT, so \trp chips 0, 1, and 4 serve one half of the PMT, and \trp chips 2, 3, and 5 serve the other.  
Note this means any given pixel on a PMT is serviced by two separate \trp chips - one chip with one channel operating in a low-gain mode (low amplification of signal), and one chip with two channels, one in a medium-gain mode and one in a high-gain mode (large amplification of signal). 
All three channels operate simultaneously and independently of each other, each recording a value proportional to the gain times the charge.
In principle, every channel on a \trp chip supports a discriminator, but only the discriminators on the high gain \trp channels are active. 
This is because the high gain channels are the most sensitive and low and medium-gain channels cannot be over threshold without their corresponding high-gain channel also being over threshold.
The channel map ties two high and medium-gain \trp chips to each low gain \trp chip.  
In order to associate data in the low gain with data in the high or medium for a given channel, the low-gain \trp for that channel must be pushed with the high gain \trp chip.  
To avoid ambiguous hit assignment for the \emph{other} channels serviced by the low-gain \trp chip that are not shared with the firing high-gain chip, the ``parallel'' high-gain chip must also be pushed.  
Thus, there are effectively two logical units, 32 channels in size, on each board that are used to fill the analog pipelines.

\subsection{Clock Ticks and Time}
\label{sec:clockticks}
Time on the FEBs is set relative to a free-running counter that counts up in clock ticks until it reaches the 32-bit unsigned integer maximum, at which point it simply rolls over to zero.  
The counters are reset to the value stored in the programming registers of the FPGA control chips when the FEBs receive a CNRST signal.  
By tuning this start value to account for signal propagation time, it is possible to synchronize the local counter values on the FEBs to within one clock tick.

After a gate-open signal, the FEB loads its Timer register value and starts a counter that counts up to the 16-bit unsigned integer maximum.  
This delay is used to synchronize with beam activity and as a "quiet" period to blank charge pulses to the CW voltage system just before SGATE (at least 15 microseconds is required).  
At the counter maximum the FEB gate opens and the time on the free-running counter is latched as our gate start time.
It is an offline software convention to subtract this time from all subsequent calculations on a board.

Because the low gain channels are shared across two high and medium gain sets of channels, when the discriminators fire charge is pushed and stored in groups of 32 channels.  
This means a hit in a given channel creates a timestamp $T_0$ for the whole \trp chip via the high gain.
That timestamp is further shared across the ``\trp Chip Group'' (0, 1, and 4 share the same base time, and 2, 3, and 5 share a separate base time).  
Subsequent hits within the 16-tick integration window have differing delay and quarter ticks.
Crossing a discriminator threshold causes all three associated \trp chips to push 16 FEB clock ticks later, regardless of subsequent hits.  

When a discriminator fires, it sets its whole \trp chip group (0+1+4 or 2+3+5) to push 20 system ticks later. 
The entire group shares the base integer time stamp for the earliest channel to fire on the group, and the initiating channel also stores a quarter-tick for fine resolution. 
The system clock continues to run and provides delay and quarter ticks to timestamp subsequent hits (prior to the push) to 1/4 of a tick accuracy in other channels (pixels) in the group.

\subsection{Dead Time}
\label{sec:deadtime}
\begin{figure}[hbtp]
  \centering 
  \includegraphics[width=1.0\textwidth]{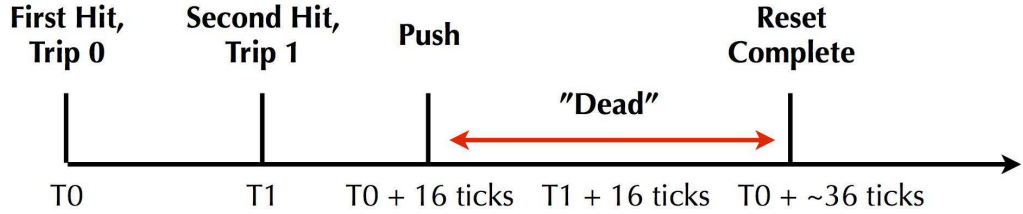}
  \caption{A specific dead-time illustration on one FEB.  Consider a hit on \trp 0.  This hit will set a timestamp, $T_0$ here, and cause a pipeline push 16 clock ticks later for \trp chips 0, 1, and 4.  A later hit at time $T_1$ above does not initiate a new pipeline push and the time is stamped relative to $T_0$.  The ADCs are effectively dead between $T_0 + 16$ and $T_0 + 36$ clock ticks.  After that point, the ADCs on those TRiPs are fully live again and new hits over threshold will be timestamped.}
  \label{fig:deadtime}
\end{figure}

\begin{figure}[hbtp]
  \centering 
  \includegraphics[width=1.0\textwidth]{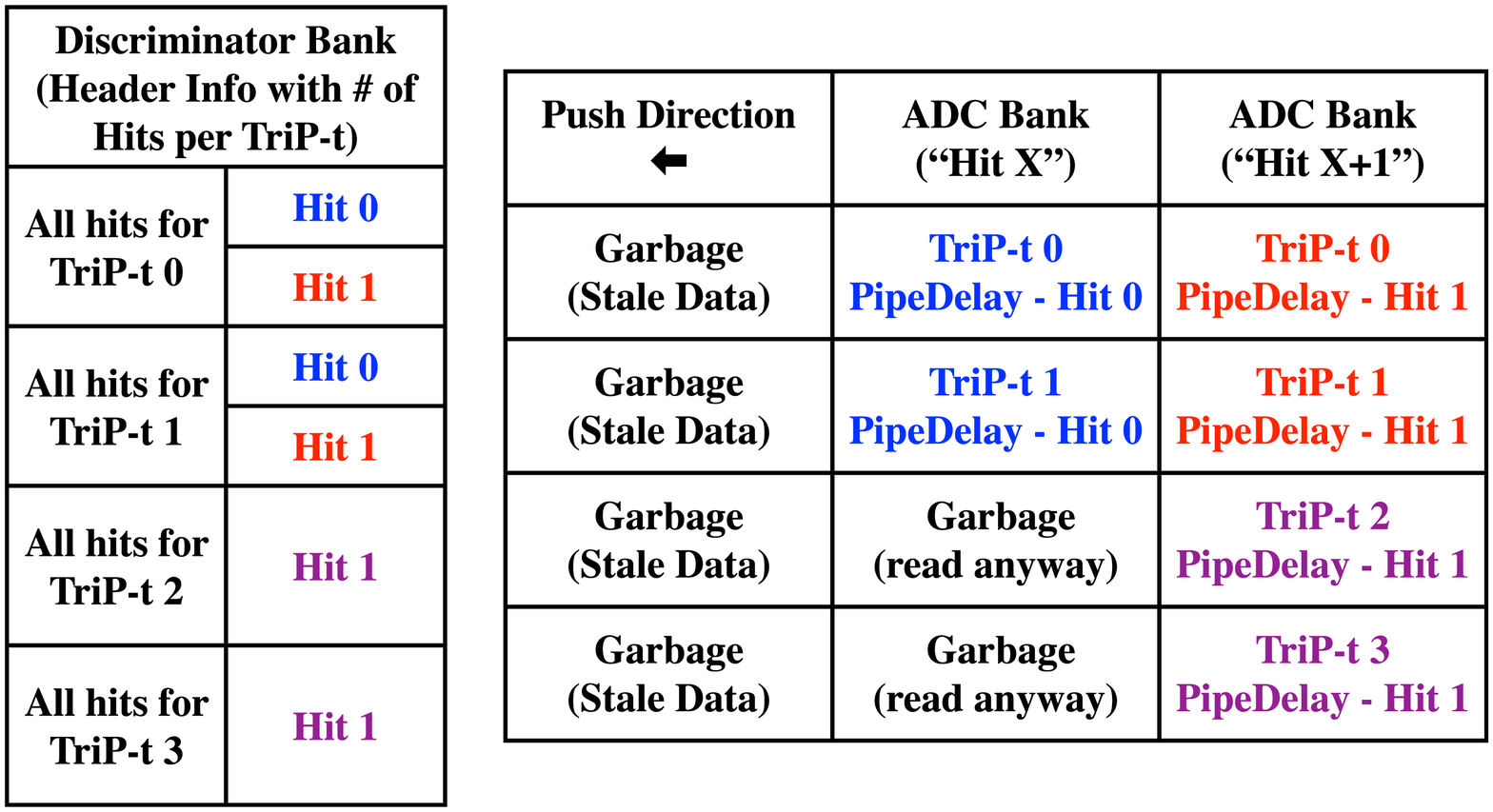}
  \caption{
  A generic example of the organization of data within the discriminator and ADC banks.  
  The columns (``Hit X'') denote hit RAM Functions - all of the data in the column is read at once and partial pipeline reads cannot be performed.
  In this case, there was a hit on either \trp 0 or 1, another independent hit on either \trp 2 or 3, and a final hit on \trp 0 or 1.  
  The ADCs cannot distinguish which \trp in the two pairs contained the initial hit for that pair. 
  Again, this diagram does not distinguish which \trp had the channel hit first, and it cannot distinguish which of the hits in the ``Hit 1'' RAM Function was earlier - the discriminator bank must be decoded for that information.  
  The end of integration ``hit'' is not included in the set of ADC banks shown here.   
  It would appear in a new column titled ``Hit 2'' and would appear in all \trp's. 
  For each FEB read out, the following RAM Functions: ReadDiscr, ReadHit0, ..., ReadHit7 are looped through.  
}
  \label{fig:banksillus}
\end{figure}

\begin{figure}[hbtp]
  \centering 
  \includegraphics[width=0.7\textwidth]{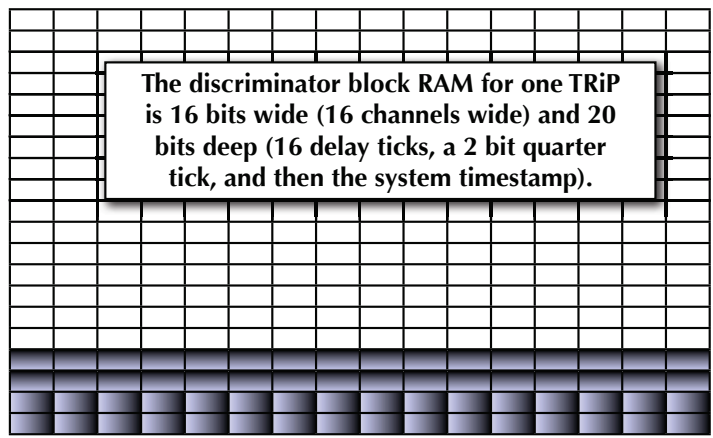}
  \caption{A visualization of the discriminator bank for a single \trp chip.  The system tick is stored in the 32 bits at the ``bottom'' of the figure.  Quarter ticks for each channel are stored in the 32 bits above that at 2 bits per channel.  Finally, the delay ticks are stored in 16 bits for each channel above the quarter ticks.}
  \label{fig:discrbank}
\end{figure}

Coupling pairs of high gain \trp chips implies dead time impacts 32 channels at a time. 
The reset after a push requires 20 clock ticks to complete. 
During the reset period, all three \trp chips that pushed are essentially dead to incoming signals. 
Hits during the reset period are un-timed and the charge recorded is either greatly attenuated or zero.
For illustration, consider Fig. \ref{fig:deadtime}.
When there is a hit on one half of a discriminator-active \trp pair at time $T_0$ and another hit at a later time $T_1$, the second hit is latched in time as long as it arrives before the push.
If the second hit on either pair of \trp chips arrives more than 16 ticks after the initiating hit, the later hit is not recorded.
Time on both \trp chips is relative to $T_0$, the first threshold cross between the two.   
Fig. \ref{fig:banksillus} shows a typical bank structure for a gate and
Fig. \ref{fig:discrbank} shows the memory structure of a discriminator bank.

There is another form of dead time due to hit saturation.  
This occurs when the number of ``pushes'' allowed by the pipeline depth are exhausted. 
In this case, charge continues to integrate, but additional hits are no longer time-stamped.

\subsection{FEB Timing Synchronization}
\label{sec:febsync}
When the CRIM sends an open-gate command (SGATE) through the CROCs to the FEBs, the actual signal arrives at different times to the different FEBs on the chain due to cable delays.  
This is accounted for in two steps:
\begin{enumerate}
\item The FEB loads the value recorded in the Timer register of the FPGA.  
By staggering the time loaded to account for the finite signal propagation time down the chain, all of the boards can synchronize the values on their local counters.
\item Once the FEBs receive an open-gate signal they begin counting up to a pre-set open-gate time value, beginning not with zero, but with the Timer value mentioned in the previous step.
This additional delay is meant to provide the PMT with time to settle into a stable high voltage in the case the Cockroft-Walton adjusted the voltage period just before the gate is opened. 
Once they reach this time, they open their integration gates.  
All the FEBs will physically open their gates at the same local time.
\end{enumerate}

\section{DAQ Software}
\label{sec:daqsw}
\subsection{Readout Control}
\label{sec:readoutcontrol}
The DAQ  is a system built out of three computers running  64-bit Scientific Linux Fermi  \cite{refslf} - one readout node for each VME crate and one master node.  
Data transfer between the readout nodes and master node is handled via the Event Transfer (ET) software package (part of CODA \cite{refcoda}), discussed further below.  
The head node initiates a run and creates a memory-mapped file to store data.  
It provides instructions to the readout nodes on when to begin the readout sequence.  
When readout begins, each of the slave nodes passes data frame-by-frame to the master for collation.  
At the end of a run, the head node converts the memory-mapped file into a flat binary and releases it.  
One feature of this approach is that if the DAQ fails during readout, the collected data are already stored on disk and are not lost.
The DAQ is written in C++ to run on Scientific Linux Fermi \cite{refslf}, but should run with minimal modifications on most major Linux distributions.  
The code is divided into four libraries:
\begin{enumerate}
\item Hardware access: Our hardware access libraries are built on top of a set of proprietary (closed-source) libraries provided by CAEN to access the V2718 VME crate controller \cite{refcaen2718} using the A2818 PCI optical bridge card \cite{refcaen2818}.  
Functions from the libraries CAEN makes freely available for this hardware on its website are used to format command and control messages to the CROCs and CRIMs as required during readout. 
\item Event structure: Another set of libraries that define the types and structures of the $\mnv$ data types. 
\item Acquisition control: These libraries and executables run the acquisition and interface with the Run Control software described in Section \ref{sec:runcontrol}.
\item Buffer management and interprocess communication (IPC): Like the hardware access libraries, the buffer management and IPC code uses external libraries.  In particular, the open-source Event Transfer (ET) system version 9.0, based on CODA, \cite{refcoda} developed at the Thomas Jefferson Lab National Accelerator Facility (JLab) \cite{refjlab} is used.
\end{enumerate}

\subsection{Run Control}
\label{sec:runcontrol}
The run control is written in Python for flexibility and platform-independence.  
There are two external Python libraries (in addition to the DAQ software, described in section \ref{sec:readoutcontrol}) needed for various components of the run control to work:
\begin{enumerate}
  \item wxPython \cite{refwxpython}, chosen as our graphics toolkit because of its ease of use and maturity;
  \item pySerial \cite{refpyserial}, which provides a convenient RS-232 interface.
\end{enumerate}

\section{Computing Architecture}
\label{sec:comparch}
\subsection{Overview of Data Handling}
\label{sec:arch_datahandlingoverview}
$\mnv$ uses two networks of server-class computers: one for physics data collection and one for data quality monitoring and run control.  
The data collection PCs are detailed in Section \ref{sec:physicalorg}.
For monitoring, $\mnv$ uses two remote clusters of computers at Fermilab: one in the Feynman Computing Center (FCC), and the other at Wilson Hall (WH).  
These systems are discussed in Sections \ref{sec:controlroom} and \ref{sec:datahandlingpaths}

Data are collected and logged underground independently of above-ground computing clusters. 
The DAQ can run without input from or a connection to any above-ground network with local storage for physics data, metadata, and logs sufficient for about a month of normal operations.
The above-ground computers manage long-term data archival, log file metadata into SAM (Sequential Access Method - a metadata database application) \cite{refsamm}, produce histograms for monitoring, and provide run control options.

\subsection{Data acquisition coordination}
\label{sec:arch_coordination}
The run control software manages connections between most of the computers used in the data acquisition and monitoring process such that they remain synchronized throughout.  
Kerberos-authenticated \cite{refkerberos} SSH tunnels are used to connect clients (i.e., user shift consoles) to a master DAQ server process and the master DAQ server to the other slave machines, as described below.

The heart of the run control is a central server process on the master DAQ node.
It maintains communication with ``dispatcher'' processes running on each of the relevant slave remote nodes so as to instruct them to start or stop their various tasks as the state of data acquisition changes.  
It is also the entry point for communication from the graphical end client used by operators to interact with the DAQ.  
This ``DAQ manager'' process can coordinate data taking irrespective of the number of clients connected; once started, it will run through to completion of the data acquisition sequence even if the controlling client is disconnected.

The DAQ manager connects to a collection of remote nodes in various locations.  
Among these are readout nodes that insert the digitized data into the data stream and the monitoring nodes that run slow monitoring processing jobs and serve the data to the operator.  

Operators control the DAQ using a graphical client (also contained in the run control package) that connects to the DAQ manager over the network using a custom thin-client TCP protocol.  
This design accommodates both the Fermilab control room and collaborator institutions located around the world.  

\section{Physical Organization of the Readout Equipment in NuMI}
\label{sec:physicalorg}
All relevant $\mnv$ electronics equipment is mounted in racks sitting on a deck beside the detector.  
All racks are equipped with smoke detection and interlocked power, with rack protection and slow monitoring built into the overall online scheme.  

\begin{figure}[hbtp]
  \centering 
  \includegraphics[width=1.0\textwidth]{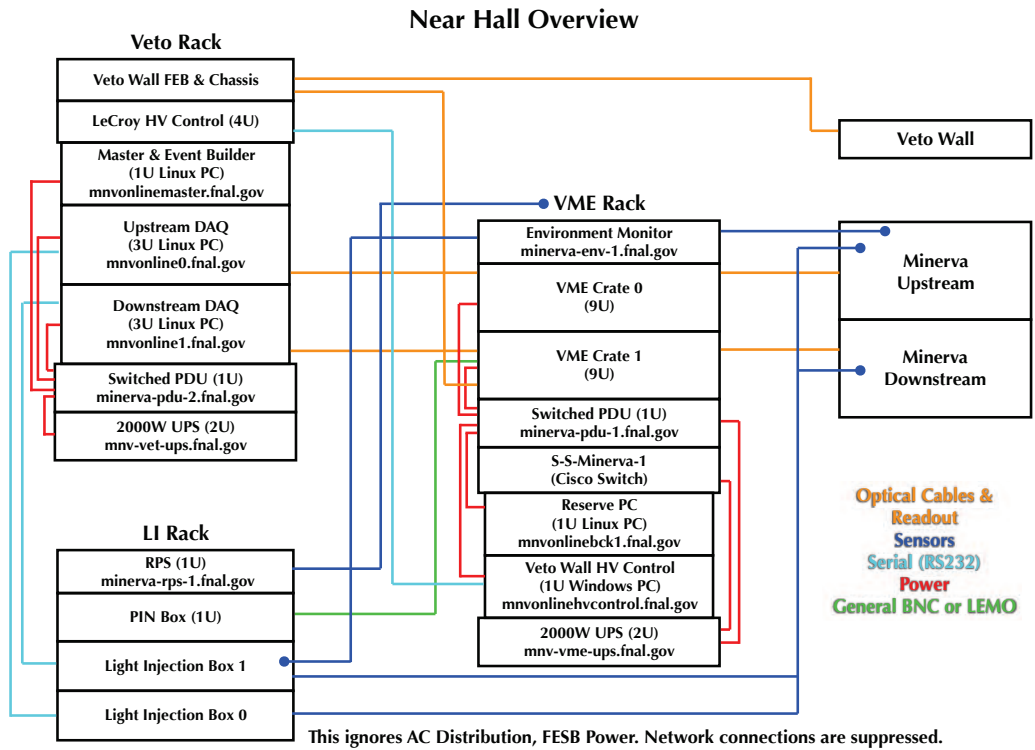}
  \caption{Connection map for the $\mnv$ readout system, ignoring AC distribution, FESB power, and safety systems.  
Network connections are also suppressed - every device with an internet name is linked to the network switch housed in the VME rack.}
  \label{fig:nearhalloverview}
\end{figure}

Racks are labeled according to their primary hardware electronics components and there are four primary racks:
\begin{enumerate}
\item The VME Rack (also sometimes called the DAQ Rack)
\item The Light Injection (LI) Rack
\item The Veto Rack
\item The Spares Rack
\end{enumerate}
Our racks all consume less than 2500W each, meeting the FNAL requirements for air cooling.
See Fig. \ref{fig:nearhalloverview} for an overview of the arrangement of and connections between the three racks.

\subsection{The VME Rack}
\label{sec:vmerack}
\begin{figure}[hbtp]
  \centering 
  \includegraphics[width=1.0\textwidth]{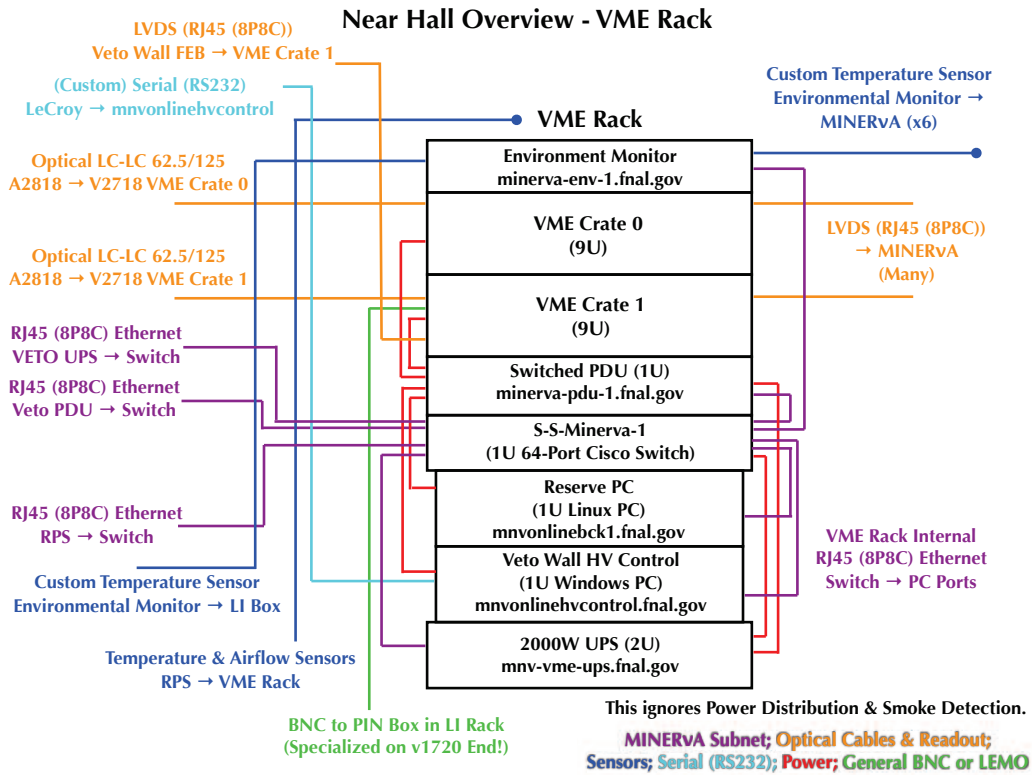}
  \caption{Connection map for the VME Rack. 
Power architecture is not shown, including the three networked fuse panels.}
  \label{fig:vmerack}
\end{figure}

\begin{figure}[hbtp]
  \centering 
  \includegraphics[width=0.8\textwidth]{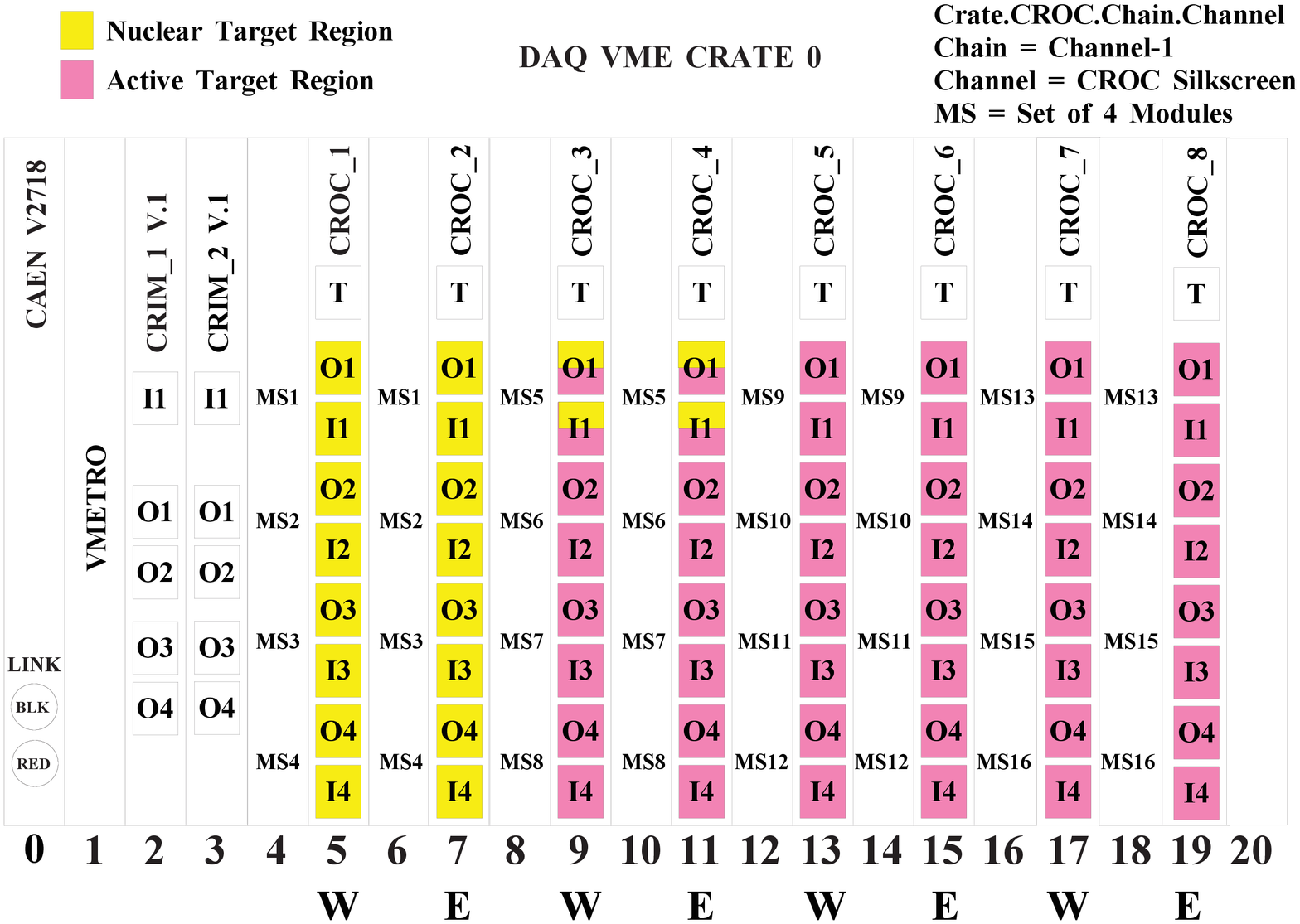}
  \caption{The connection and layout diagram for VME Crate 0.}
  \label{fig:vmecrate0}
\end{figure}

\begin{figure}[hbtp]
  \centering 
  \includegraphics[width=0.8\textwidth]{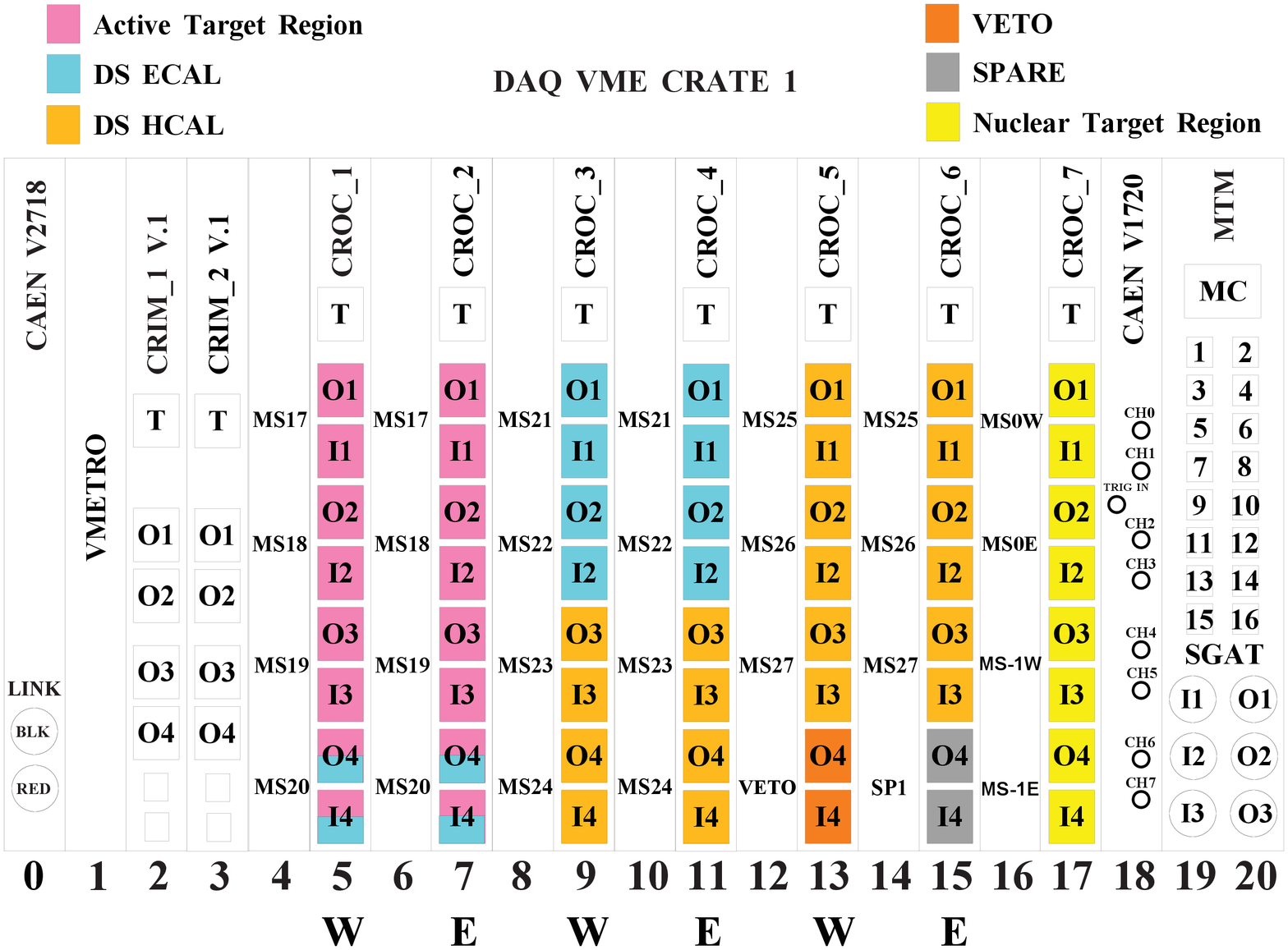}
  \caption{The connection and layout diagram for VME Crate 1. \emph{The V1720 CAEN Waveform Digitizer is not currently installed.}}
  \label{fig:vmecrate1}
\end{figure}

The VME Rack contains the two VME crates that hold the $\mnv$ readout electronics.  
It additionally mounts a pair of PCs and network support components.  
Access control for devices that cannot be configured to use Kerberos is provided by a very restrictive ACL (Access Control List).  
See Fig. \ref{fig:vmerack} for the rack layout and connection types.
Insofar as the readout architecture is concerned, it contains (roughly in mechanical installation order from top to bottom):
\begin{enumerate}
\item A custom Temperature and Environment Monitor (TEM) device (labeled the ``Slow Monitor'') that supports eight temperature probes that may be deployed throughout the racks and over the detector surface.  
The device also logs humidity, pressure, and temperature at the end of six cabled sensors deployed on the detector.   
\item Two CAEN VME 8011 crates \cite{refcaen8011}.  
The crates are indexed as 0 and 1.  
Crate 0 contains a controller module, two CRIM modules, and eight CROC modules.  
Crate 1 contains a controller module, two CRIM modules, seven CROC modules, and a MvTM module.  
See Figs. \ref{fig:vmecrate0} and \ref{fig:vmecrate1} for the VME module assignment in Crates 0 and 1 respectively.
\item An APC 120VAC Masterswitch III Switched Rack PDU (Power Distribution Unit), Model \# AP7900.  
This PDU is a networked device that provides remote power control for the VME crates and all PCs in the rack. 
\item A 48 port Cisco C2960G switch, designated S-S-MINERVA-1.  \cite{refswitch}
\item Three 48VDC power supplies to power the FEB's (not pictured in the figures).
\item Three networked fuse panels for the 48VDC power supplies, networked to S-S-MINERVA-1. 
These fuse panels host a web interface with self-documented power control of the FESBs mounted on the $\mnv$ detector (not pictured in the figures). 
\item A 1U Linux PC serving as a back-up for the master DAQ node in the Veto Rack (see Section \ref{sec:vetorack}).
\item A 1U Windows PC running control and monitoring software for the LeCroy High Voltage Power System (HVPS) located in the Veto Rack. \emph{Not currently installed. Space, power, and networking resources are reserved.}
\item An  APC Smart-UPS 2200VA (2000W Un-interruptible Power Supply) supporting the switch and PCs.  
The primary function of this UPS is power filtering. 
In the event of a full power failure, the system will remain live for more than thirty minutes.
\end{enumerate}

\subsection{The LI Rack}
\label{sec:lirack}
\begin{figure}[hbtp]
  \centering 
  \includegraphics[width=1.0\textwidth]{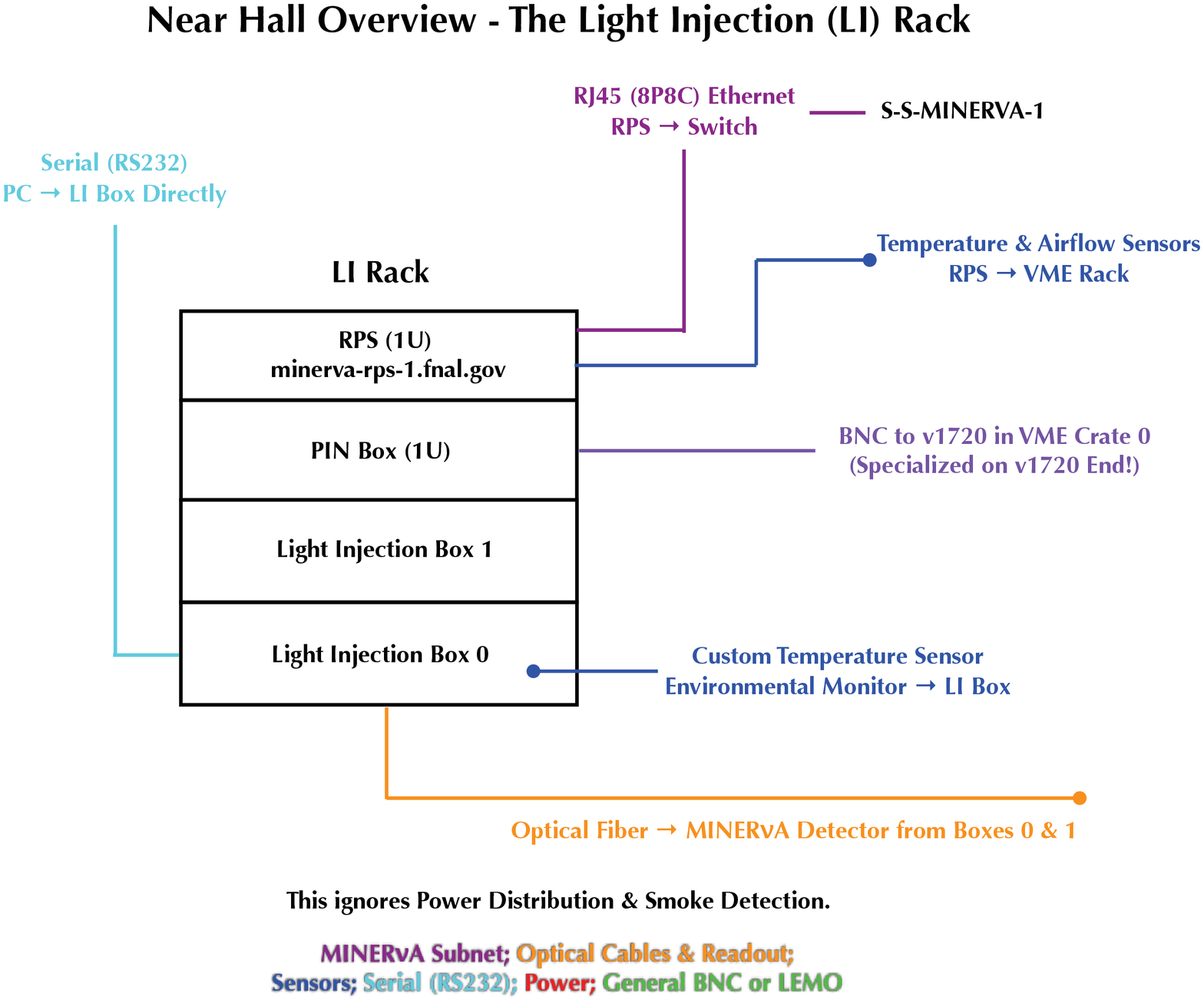}
  \caption{Connection map for the LI Rack.}
  \label{fig:lirack}
\end{figure}

The LI rack contains the $\mnv$ custom-built Light Injection (LI) boxes \cite{refli}.  
The rack is intentionally kept sparse to provide sufficient space to route and store the optical fibers that connect from the back of the LI boxes to the PMTs mounted on top of $\mnv$.  
See Fig. \ref{fig:lirack} for the rack layout and connection types.
For our purposes, it contains (roughly in mechanical installation order from top to bottom):
\begin{enumerate}
\item A BiRa Rack Protection System (RPS).  
The RPS deploys a variety of monitoring cables into the VME Rack: voltage-monitoring cables to the VME crates, and airflow and temperature sensors.  
The RPS is connected to the S-S-MINERVA-1 switch located in the VME rack by ethernet.
\item A PIN diode box.  
The PIN is optically connected to the LI box to monitor the stability of the LED light output.  The PIN box is additionally connected via BNC and specialized cables to a CAEN V1720 waveform digitizer \cite{refcaen1720} mounted in VME Crate 0 in the VME Rack.
\item The primary LI box.  
This box contains control electronics and LEDs that service optical connectors on the back of the box.  From those connectors light is piped via optical fiber to the PMT boxes mounted on top of the detector.  
It is interfaced via a 9-pin serial RS-232 socket on the front of the box.  
This serial cable is connected to the $\mnv$ DAQ PC cluster for configuration commands.  
It additionally accepts a LEMO connection for the external trigger signal.  
\item A secondary (smaller) LI box to service the most forward detector modules. It uses the same electronics and connections as the larger box, but mounts fewer LEDs.
\end{enumerate}

\subsection{The Veto Rack}
\label{sec:vetorack}
\begin{figure}[hbtp]
  \centering 
  \includegraphics[width=1.0\textwidth]{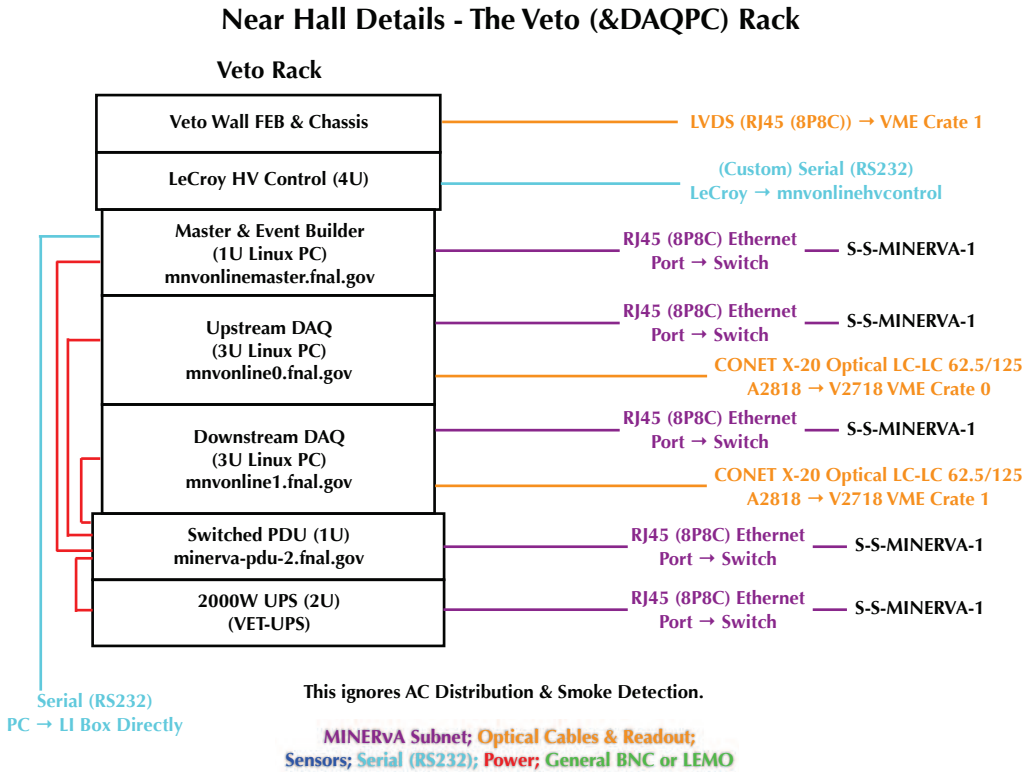}
  \caption{Connection map for the Veto Rack.}
  \label{fig:vetorack}
\end{figure}

The Veto rack derives its name from the LeCroy High Voltage Power System mounted there that controls the high voltage on the resistor-base PMTs deployed to readout the "Veto Wall" sitting in front of the $\mnv$ detector.  
It also contains the main readout PCs for the DAQ system.  
See Fig. \ref{fig:vetorack} for the rack layout and connection types.
For our purposes, its important contents are:
\begin{enumerate}
\item Two $\mnv$ FEBs attached to a special custom breakout board in a 2U chassis.  
The FEBs are standard $\mnv$ boards, but the breakout board is required to interface the single-anode PMTs (resistive base) used to monitor the scintillator panels in the veto wall.
\item A LeCroy Model HV4032A High Voltage Power System (HVPS).  
The HVPS connects to the readout PC located in the VME Rack via a special custom serial cable. 
It also connects to the Veto Wall PMTs.
\item A 1U Linux PC running the DAQ Master software.  This PC is identical to the logger PC in the VME Rack.  The DAQ Master node connects to the Slave Readout nodes via direct ethernet.   
\item Two 3U Linux PCs serving as DAQ Readout Slave nodes.  
Each PC mounts a CAEN a2818 PCI Optical Bridge card \cite{refcaen2818} and interfaces with one of the VME crates located in the VME Rack.  
\item An APC 120VAC Masterswitch III Switched Rack PDU (Power Distribution Unit), Model \# AP7900.  
This PDU is a networked device that provides remote power control for the VME crates and all PCs in the rack.  
\item An  APC Smart-UPS 2200VA (2000W Un-interruptible Power Supply) supporting the PCs.  
The primary function of this UPS is power filtering.  
In the event of a full power failure, the system will remain live for over thirty minutes.
\end{enumerate}

\subsection{The Spares Rack}
\label{sec:sparesrack}
The final rack in the $\mnv$ electronics installation sits across the hall from the detector on the West side of the enclosure. 
It contains a number of spare components:
\begin{enumerate}
\item A 3U Linux PC configured to serve as a spare readout node.
\item A 48 port Cisco C2960G switch, designated S-S-MINERVA-2. \cite{refswitch}
\item A 1U Linux PC running logging software for the slow monitoring devices.  
This machine also serves as the head node for a pool of servers running a Condor \cite{refcondor} queue for batch processing jobs used to support monitoring activities in the control room.
\item Spare AC distribution and smoke protection units.
\end{enumerate}

\subsection{Power Infrastructure}
\label{sec:power}
\subsubsection{DC Power Distribution}
\label{sec:dc_power_dist}
The large number of Front End boards and their location atop the detector makes conventional low-voltage power distribution impractical.  
A distributed power architecture is employed where 48 VDC power equipment is located in the electronics racks and intermediate low voltage regulation is performed on the detector near the FEBs.  
Finally, low-noise linear regulators on each FEB produce the final 3.3 V power required.

Each FEB readout chain (up to ten FEBs) is powered by a Front End Support Board (FESB) which contains an isolated DC-DC converter to step down 48 VDC to approximately 4 VDC at 20 A.  
FESBs are passively cooled and use the FEB/PMT steel mounting bracket as a heat sink.  

Three 48 VDC power supplies operate in parallel and support ``hot swap'' and dynamic load sharing.  
Two supplies are required to meet the design specification of 3000W.  
The extra power supply allows for ``N+1'' redundant operation, which makes it simple to remove and replace a supply without interruption.  
Each power supply is rated for 1900 W and provides an interface by which the health of the supply may be monitored remotely (through the Rack Protection System (RPS) interface board and Environmental Monitor).

Through the Fuse Chassis the main 48 VDC 120 A bus is fanned out into six 20 A connections.  
Two of these 20 A connections feed a Network Fuse Panel (NFP), which fans out into twenty 2 A busses going to the FESBs.  
Each NFP offers independent control and monitoring of the FESBs through a web interface.  

\begin{figure}[hbtp]
  \centering 
  \includegraphics[width=1.0\textwidth]{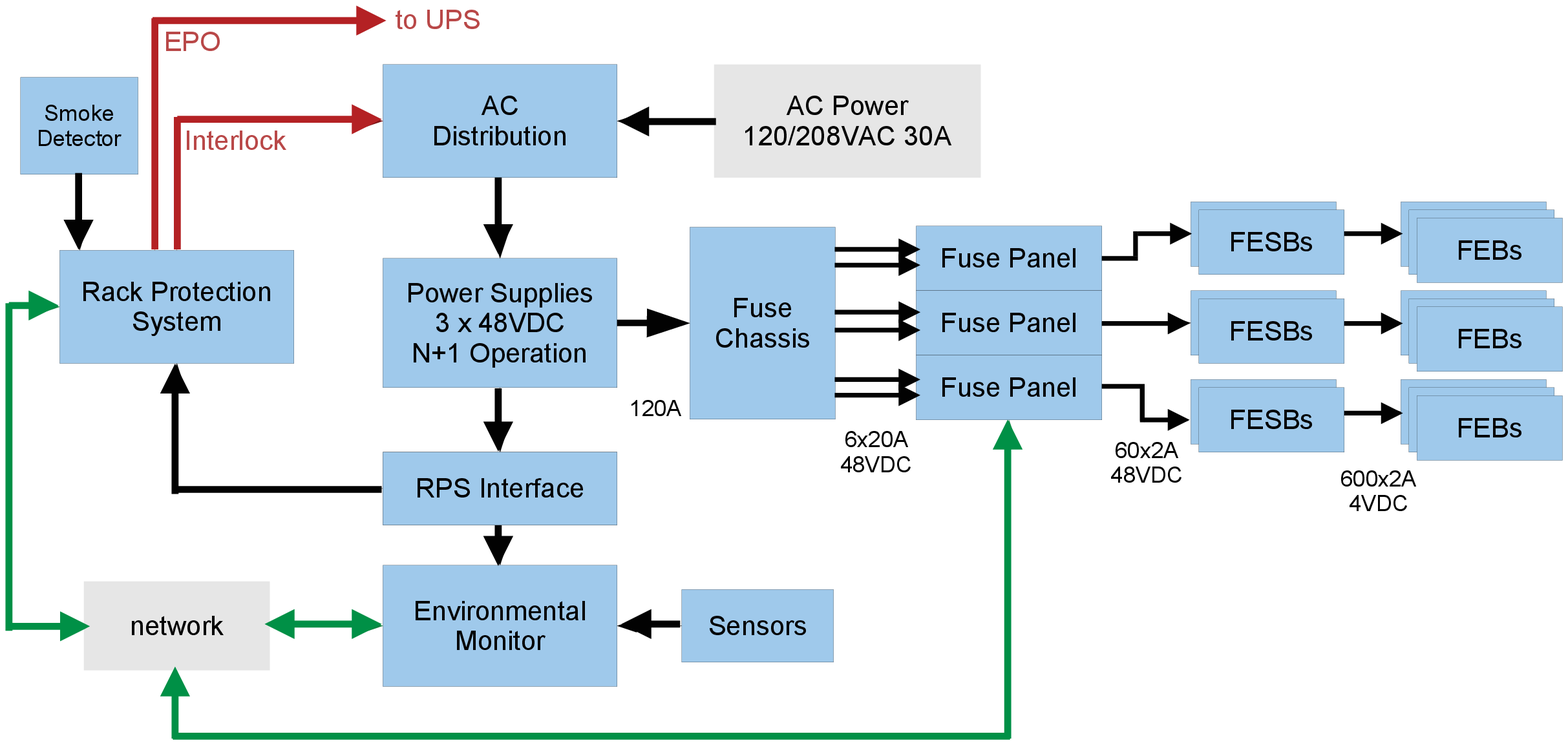}
  \caption{Power distribution for $\mnv$.}
  \label{fig:powerfig1}
\end{figure}

The actual power supply consumption is 1500 W when the FEBs are idle.  
The redundant ``N+1'' power supply configuration has been  robust and reliable and has operated continuously without problems since commissioning in 2008.
See Fig. \ref{fig:powerfig1} for a schematic.

\subsubsection{Rack Protection System and AC Power Distribution}
\label{rps_ac_power_dist}
Rack protection is provided by the BiRa Systems RPS unit model 8884 \cite{refbira}, which monitors a smoke detector in each of the three racks on the platform.  
If smoke is detected, the RPS unit drops the interlock signal, disabling the AC distribution chassis.  
The RPS also generates an emergency power output (EPO) signal that immediately disconnects the internal batteries on the UPS.

The Environmental Monitor reads various temperature sensors located around the detector. 
In addition, this monitor records and logs humidity and ambient light conditions.  
It is accessed through a web and FTP interface on the DAQ network.

\section{Interface to the $\mnv$ Control Room}
\label{sec:controlroom}
\subsection{The $\mnv$ Control Room}
\label{sec:mnvcontrolroom}
\begin{figure}[hbtp]
  \centering 
  \includegraphics[width=0.8\textwidth]{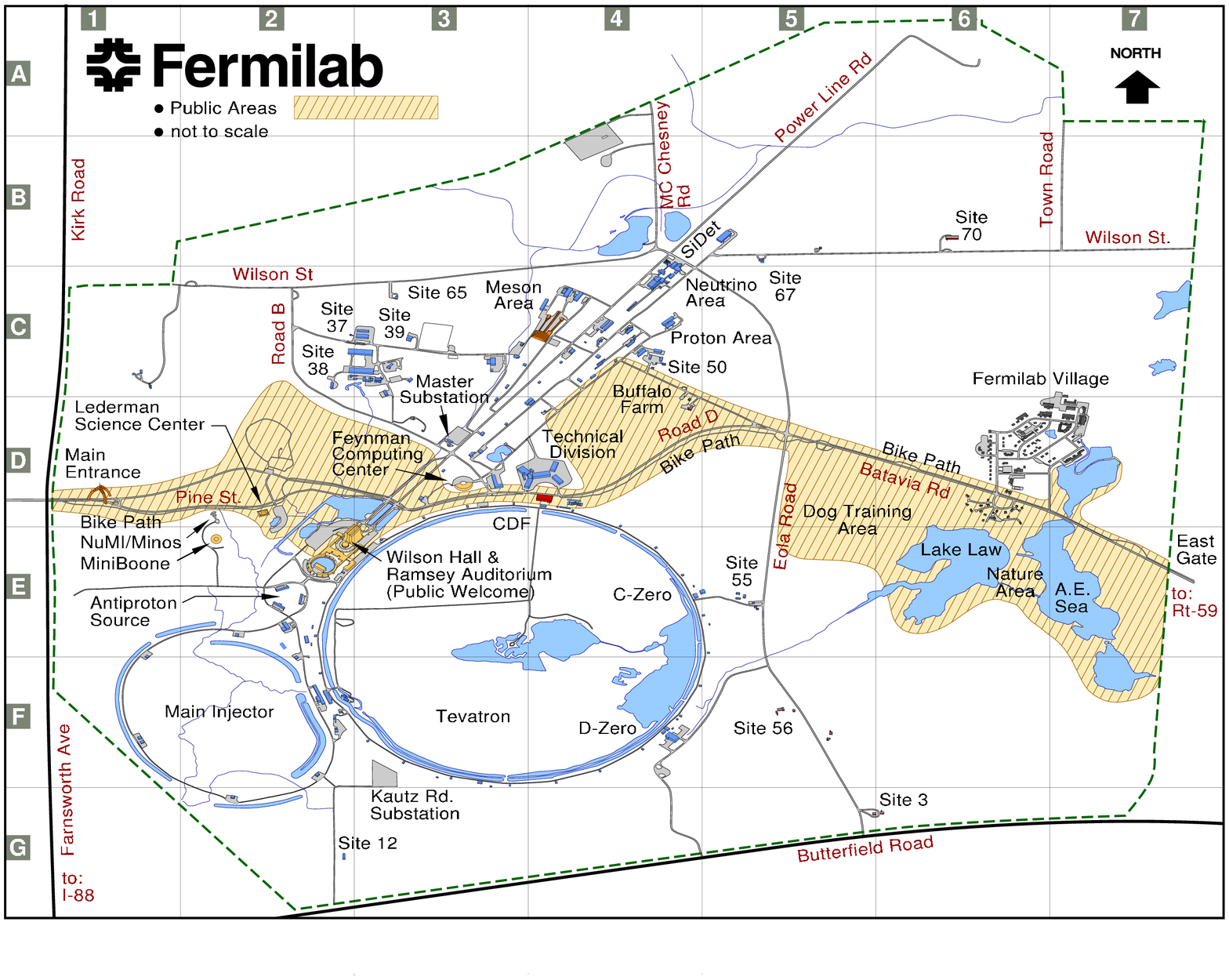}
  \caption{Buildings and sites at Fermilab.}
  \label{fig:sitemap}
\end{figure}

The $\mnv$ control room is located in Wilson Hall at FNAL, in the north-west corner room on the 12th floor (WH12).  See Fig. \ref{fig:sitemap} for a map of building locations at Fermilab.
$\mnv$ services our control room with a total of eight PCs, with four in the Control Room and four in the Feynman Computing Center (FCC) at FNAL.  
The PCs kept in FCC run more processor-intensive tasks using a private Condor batch queue system \cite{refcondor}.  

\subsection{Controlling the DAQ System}
\label{sec:iface_controlDAQ}
As noted above in section \ref{sec:arch_coordination}, the DAQ system is controlled via a graphical thin client delivered with the run control software.  
The client runs locally on the operator's terminal. 
Control of the DAQ is via a custom TCP thin-client protocol that is transmitted over a Kerberos-enabled SSH tunnel to the master DAQ node.  
Inbound connections are filtered both via the firewall provided by the NuMI router and by an access list of Kerberos principals allowed to log into the machine.

Commands sent via a successful tunnel to the DAQ manager process (running on the DAQ master node) are interpreted based on a notion of client ``control'' of the DAQ.  
At most one (though possibly zero) client(s) can be in control of the DAQ at any given time; only commands that originate from a node which is currently ``in control'' are ultimately acted upon.  
An unlimited \footnote{Unlimited at least in principle.  Network bandwidth and CPU time saturation, of course, will create an operational ceiling, though the actual maximum will depend on the environmental conditions.} number of clients are allowed to connect as observers.  
Any observer is allowed to request control at any time; however, if another client is in control at the time the request is made, the client in control is given a configurable length of time to veto the request or approve transfer of control.  
In addition, the run control package contains a client manager program that can revoke or assign control to or from any client that is currently connected after proper credentials are submitted, currently a password.

Once connected and in control, the operator is presented with a graphical interface containing details about the current data acquisition sequence and status as well as  controls allowing him/her to start, manually advance, or stop data acquisition.  
When the DAQ is stopped the operator can also adjust a handful of simple configuration options pertaining to the data acquisition sequence he/she would like to begin.  
The bulk of the configuration operations, however, is accessible only to experts directly on the DAQ head node so as to avoid accidental misconfiguration.

Once initiated, data acquisition requires no intervention from the operator.  
The run control system can even operate unsupervised by any clients at all (as would happen, for example, in the case where the network connection to a remote site is interrupted).  
Under these circumstances warnings about the system state are queued and delivered in bulk to every client who reconnects until they are acknowledged.

\section{Data Handling Paths}
\label{sec:datahandlingpaths}

The $\mnv$ electronics are capable of monitoring a wide variety of configuration states: high voltage on the (Cockcroft-Walton) PMTs deployed on the detector, whether individual boards are dead or live, and the temperature of the FEBs.  
A separate system monitors the resistive base PMTs used for the $\mnv$ Veto Wall.  
For cost-saving reasons $\mnv$ uses the same communications chain for electronics configuration and slow controls as it does for data handling.  
As a consequence, these monitoring data are packaged with the physics data directly and exist as part of that stream.

For monitoring purposes, the $\mnv$ near-online, or ``Nearline,'' cluster runs an ET client that is notified about the beginning and end of runs.  
During a run, it reads data from the DAQ head node frame-by-frame.  
When it completes an event (signified by reading the DAQ event header frame) it converts the binary data into $\mnv$ ``RawDigits'' (the most basic analysis data format) for monitoring. 
Other tasks running on the Nearline cluster decode the RawDigits and process them through the $\mnv$ software framework - producing monitoring quantities ranging from high voltage status to event displays. 
These are published to control room PCs for experimenters to study.

\begin{figure}[hbtp]
  \centering 
  \includegraphics[width=0.8\textwidth]{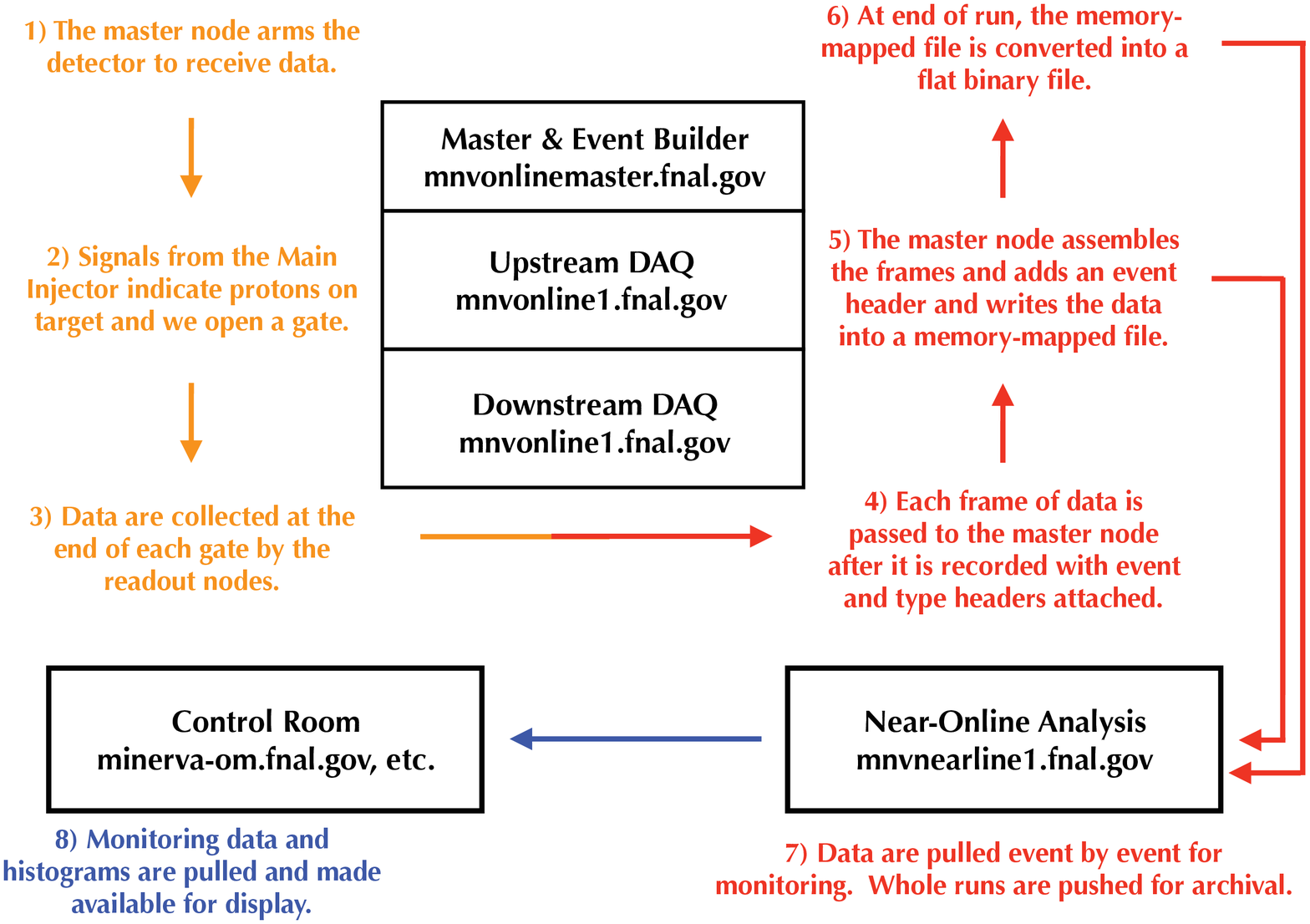}
  \caption{Flow-chart diagram for physics data through the $\mnv$ acquisition chain.  Not shown are the archival data handling beyond the near-online cluster or the metadata handling.}
  \label{fig:datflow}
\end{figure}

After each run a copy of the raw binary data file is pushed to the disk and tape for permanent storage. 
See Fig. \ref{fig:datflow} for a flowchart of the scheme.
A long-term rolling copy of recent raw data is kept on the DAQ system itself (extending back approximately two months). 
With the raw data the DAQ additionally pushes metadata about the run for entry into SAM. 
Finally, a ``keep-up'' processing that uses raw data copied directly from the DAQ (not from the Nearline system) is also run to produce the RawDigits of record.
In the end, there are three copies of the ``raw'' data at the end of this process: a copy of the raw data on the DAQ storage disks, a copy of the raw data in permanent storage, and a copy of the RawDigits (raw data reformatted for $\mnv$ software consumption) for general use.

\section{Summary and Performance}

\begin{figure}[hbtp]
  \centering 
  \includegraphics[width=0.9\textwidth]{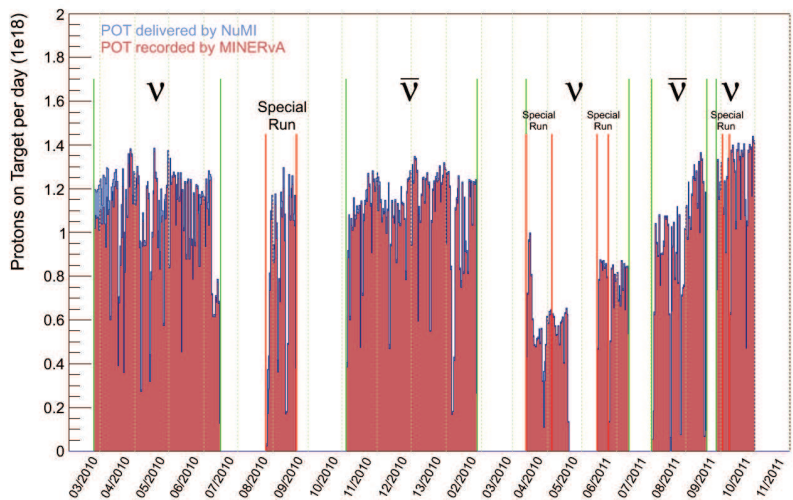}
  \caption{Protons on target delivered and recorded for $\mnv$.}
  \label{fig:potsumm}
\end{figure}

$\mnv$ began its official physics run on March 23, 2011 with the DAQ system described above. Since that time, live-time for the whole period was 96.4\% (not counting beam down-times). See Fig. \ref{fig:potsumm} for a summary of the protons on target delivered and recorded for $\mnv$. During normal operations the live-time is over 99\% and inefficiencies are almost entirely due to change-over between runs. Our integrate live-time is most impacted by electronics failures and simple operator error or inattention. There are ideas for removing losses due to run change-overs (pre-starting the subsequent run and putting it into a wait state to speed up the change) and our run control system is constantly being improved to minimize the opportunity for experimenters to make mistakes when running the system while still providing flexible operational control.





\bibliographystyle{model3a-num-names}
\bibliography{minervadaq}







\end{document}